\documentclass[12pt]{article}
\usepackage{amssymb}
\usepackage{graphicx}
\begin{document}
\newcommand{\beq}{\begin{equation}}
\newcommand{\eeq}{\end{equation}}
\newcommand{\beqa}{\begin{eqnarray}}
\newcommand{\eeqa}{\end{eqnarray}}
\newcommand{\beqar}{\begin{eqnarray*}}
\newcommand{\eeqar}{\end{eqnarray*}}
\newcommand{\al}{\alpha}
\newcommand{\be}{\beta}
\newcommand{\del}{\delta}
\newcommand{\D}{\Delta}
\newcommand{\eps}{\epsilon}
\newcommand{\ga}{\gamma}
\newcommand{\Ga}{\Gamma}
\newcommand{\ka}{\kappa}
\newcommand{\nn}{\nonumber}
\newcommand{\inn}{\!\cdot\!}
\newcommand{\h}{\eta}
\newcommand{\ii}{\iota}
\newcommand{\kk}{\varphi}
\newcommand\F{{}_3F_2}
\newcommand{\la}{\lambda}
\newcommand{\La}{\Lambda}
\newcommand{\na}{\prt}
\newcommand{\Om}{\Omega}
\newcommand{\om}{\omega}
\newcommand{\p}{\Phi}
\newcommand{\sig}{\sigma}
\renewcommand{\t}{\theta}
\newcommand{\z}{\zeta}
\newcommand{\ssc}{\scriptscriptstyle}
\newcommand{\eg}{{\it e.g.,}\ }
\newcommand{\ie}{{\it i.e.,}\ }
\newcommand{\labell}[1]{\label{#1}} 
\newcommand{\reef}[1]{(\ref{#1})}
\newcommand\prt{\partial}
\newcommand\veps{\varepsilon}
\newcommand{\pol}{\varepsilon}
\newcommand\vp{\varphi}
\newcommand\ls{\ell_s}
\newcommand\cF{{\cal F}}
\newcommand\cA{{\cal A}}
\newcommand\cS{{\cal S}}
\newcommand\cT{{\cal T}}
\newcommand\cV{{\cal V}}
\newcommand\cL{{\cal L}}
\newcommand\cM{{\cal M}}
\newcommand\cN{{\cal N}}
\newcommand\cG{{\cal G}}
\newcommand\cK{{\cal K}}
\newcommand\cH{{\cal H}}
\newcommand\cI{{\cal I}}
\newcommand\cJ{{\cal J}}
\newcommand\cl{{\iota}}
\newcommand\cP{{\cal P}}
\newcommand\cQ{{\cal Q}}
\newcommand\cg{{\it g}}
\newcommand\cR{{\cal R}}
\newcommand\cB{{\cal B}}
\newcommand\cO{{\cal O}}
\newcommand\tcO{{\tilde {{\cal O}}}}
\newcommand\bz{\bar{z}}
\newcommand\bb{\bar{b}}
\newcommand\ba{\bar{a}}
\newcommand\bg{\bar{g}}
\newcommand\bc{\bar{c}}
\newcommand\bw{\bar{w}}
\newcommand\bX{\bar{X}}
\newcommand\bK{\bar{K}}
\newcommand\bA{\bar{A}}
\newcommand\bZ{\bar{Z}}
\newcommand\bxi{\bar{\xi}}
\newcommand\bphi{\bar{\phi}}
\newcommand\bpsi{\bar{\psi}}
\newcommand\bprt{\bar{\prt}}
\newcommand\bet{\bar{\eta}}
\newcommand\btau{\bar{\tau}}
\newcommand\hF{\hat{F}}
\newcommand\hA{\hat{A}}
\newcommand\hT{\hat{T}}
\newcommand\htau{\hat{\tau}}
\newcommand\hD{\hat{D}}
\newcommand\hf{\hat{f}}
\newcommand\hK{\hat{K}}
\newcommand\hg{\hat{g}}
\newcommand\hp{\hat{\Phi}}
\newcommand\hi{\hat{i}}
\newcommand\ha{\hat{a}}
\newcommand\hb{\hat{b}}
\newcommand\hQ{\hat{Q}}
\newcommand\hP{\hat{\Phi}}
\newcommand\hS{\hat{S}}
\newcommand\hX{\hat{X}}
\newcommand\tL{\tilde{\cal L}}
\newcommand\hL{\hat{\cal L}}
\newcommand\MZ{\mathbb{Z}}
\newcommand\tG{{\tilde G}}
\newcommand\tg{{\tilde g}}
\newcommand\tphi{{\widetilde \Phi}}
\newcommand\tPhi{{\widetilde \Phi}}
\newcommand\te{{\tilde e}}
\newcommand\tk{{\tilde k}}
\newcommand\tf{{\tilde f}}
\newcommand\ta{{\tilde a}}
\newcommand\tb{{\tilde b}}
\newcommand\tc{{\tilde c}}
\newcommand\td{{\tilde d}}
\newcommand\tm{{\tilde m}}
\newcommand\tmu{{\tilde \mu}}
\newcommand\tnu{{\tilde \nu}}
\newcommand\talpha{{\tilde \alpha}}
\newcommand\tbeta{{\tilde \beta}}
\newcommand\trho{{\tilde \rho}}
 \newcommand\tR{{\tilde R}}
\newcommand\teta{{\tilde \eta}}
\newcommand\tF{{\widetilde F}}
\newcommand\tK{{\widetilde K}}
\newcommand\tE{{\tilde E}}
\newcommand\tpsi{{\tilde \psi}}
\newcommand\tX{{\widetilde X}}
\newcommand\tD{{\widetilde D}}
\newcommand\tO{{\widetilde O}}
\newcommand\tS{{\tilde S}}
\newcommand\tB{{\tilde B}}
\newcommand\tA{{\widetilde A}}
\newcommand\tT{{\widetilde T}}
\newcommand\tC{{\widetilde C}}
\newcommand\tV{{\widetilde V}}
\newcommand\thF{{\widetilde {\hat {F}}}}
\newcommand\Tr{{\rm Tr}}
\newcommand\tr{{\rm tr}}
\newcommand\STr{{\rm STr}}
\newcommand\hR{\hat{R}}
\newcommand\M[2]{M^{#1}{}_{#2}}

\newcommand\bS{\textbf{ S}}
\newcommand\bI{\textbf{ I}}
\newcommand\bJ{\textbf{ J}}

\begin{titlepage}
\begin{center}

\vskip 2 cm
{\LARGE \bf  Exploring $\beta$-symmetry in heterotic theory \\ \vskip .2 cm at order $\alpha'$ with boundary term
 }\\
\vskip 1.25 cm
  Mohammad R. Garousi\footnote{garousi@um.ac.ir}

\vskip 1 cm
{{\it Department of Physics, Faculty of Science, Ferdowsi University of Mashhad\\}{\it P.O. Box 1436, Mashhad, Iran}\\}
\vskip .1 cm
 \end{center}

\begin{abstract}

This paper investigates the $\beta$-symmetry of the heterotic string theory at order $\alpha'$ in the context of open spacetime manifolds. Our analysis reveals that the parity-odd component of the effective action at this order remains invariant under $\beta$-transformations. Furthermore, we demonstrate that the corresponding $\beta$-transformations leave the Gibbons-Hawking boundary term invariant. 
\end{abstract}

\end{titlepage}

\section{Introduction}

It is well-known that the classical effective action of bosonic and heterotic string theories, obtained through Kaluza-Klein (KK) reduction on a torus $T^{(d)}$ with $1\leq d \leq D$, is invariant under rigid $O(d,d)$ transformations at all orders of $\alpha'$ \cite{Sen:1991zi,Hohm:2014sxa}. This symmetry has been successfully employed in several studies, including \cite{Garousi:2019mca,Garousi:2019xlf,Garousi:2020gio,Garousi:2021yyd,Garousi:2021cfc}, where the $d=1$ case was considered to derive the effective action of string theory in both open and closed spacetime manifolds.

The $O(d,d)$ group for $d>1$ includes geometrical subgroups consisting of rigid diffeomorphisms and shifts on the $B$-field, as well as non-geometrical $\beta$-transformations given by
\beqa
\delta E_{\mu\nu} &=& -E_{\mu\rho}\beta^{\rho\sigma}E_{\sigma\nu}, \nonumber \\
\delta\Phi &=& \frac{1}{2}E_{\mu\nu}\beta^{\mu\nu}, \labell{b-trans}
\eeqa
where $\Phi$ is the dilaton, $E_{\mu\nu}=G_{\mu\nu}+B_{\mu\nu}$, and $\beta^{\mu\nu}$ is a constant antisymmetric bi-vector \cite{Baron:2022but}.
At the critical dimension $D$, a diffeomorphism-invariant and $B$-field gauge-invariant effective action is naturally invariant under the geometrical subgroups of the $O(D,D)$ group with non-constant parameters. The invariance of the $D$-dimensional covariant action under $\beta$-transformations would indicate that the effective action is invariant under the rigid $O(D,D)$ transformations without relying on KK reduction.

In a recent paper \cite{Baron:2022but}, it was suggested that the effective actions at the critical dimension $D$ may be invariant under $\beta$-transformations if a constraint is imposed on the partial derivatives, given by
\begin{equation}
\beta^{\mu\nu}\partial_{\nu}(\cdots)=0. \labell{b-cons}
\end{equation}
Using the frame formalism, it has been shown that the effective action at the leading order of $\alpha'$, given by
\begin{equation}
\bS^0= -\frac{2}{\kappa^2}\int d^{D} x\sqrt{-G} e^{-2\Phi} \left( R - 4\nabla_{\mu}\Phi \nabla^{\mu}\Phi+4\nabla_\mu\nabla^\mu\Phi-\frac{1}{12}H^2\right), \labell{S0bf}
\end{equation}
where $H=dB$, is indeed invariant under $\beta$-transformations.  In the bosonic string theory, $D=26$, and for the superstring theory and the heterotic string theory, $D=10$. The existence of this symmetry at the supergravity level was already noted in \cite{Andriot:2014uda}. In \cite{Baron:2022but}, it was shown that the transformation of the effective action at order $\alpha'$ in the two-parameter generalized Bergshoeff-de Roo scheme \cite{Marques:2015vua} results in three- and four-flux terms. It was demonstrated that these terms can be cancelled by deforming the $\beta$-transformations. It is important to note that there are no residual total derivative terms in proving the $\beta$-symmetry of the above actions.

The effective action \reef{S0bf} is the correct effective action of string theory for closed spacetime manifolds. However, for open spacetime manifolds, this action does not satisfy the least action principle. In fact, the least action principle for the metric indicates that there must be the Gibbons-Hawking boundary term \cite{Gibbons:1976ue}. Assuming that the effective actions at the critical dimension are background-independent \cite{Garousi:2022ovo}, one would expect that the global $\beta$-symmetry is the symmetry of the more general spacetime backgrounds with boundary. In \cite{Garousi:2022qmk}, it was shown that the effective action for open spacetime manifolds at the leading order of $\alpha'$, given by $\bS^{(0)}+\partial \!\!\bS^{(0)}$, is indeed invariant under $\beta$-transformations. This action is expressed as
\beqa
\bS^{(0)}\!+\!\partial \!\!\bS^{(0)} \!= \!-\frac{2}{\kappa^2}\left[\int d^{D}x \sqrt{-G} e^{-2\Phi} \left( R + 4\nabla_{\mu}\Phi \nabla^{\mu}\Phi-\frac{1}{12}H^2\right)\!\!+ \!\!2\int d^{D-1}\sigma\sqrt{|g|}  e^{-2\Phi}K\right], \labell{baction}
\eeqa
where the last term is the Gibbons-Hawking boundary term \cite{Gibbons:1976ue}. In the given term, $K$ represents the extrinsic curvature of the boundary. This curvature is defined as $
K_{\mu\nu}=\nabla_{\mu}n_{\nu}\pm n_{\mu}n^{\rho}\nabla_{\rho}n_{\nu}
$, where $n_\mu$ refers to the outward unit vector normal to the boundary. The choice of minus (plus) sign corresponds to a timelike (spacelike) boundary.  In proving the $\beta$-symmetry of the above bulk action, one encounters residual total derivative terms that can be cancelled with the $\beta$-transformation of the Gibbons-Hawking boundary term.
If the $\beta$-symmetry is to be preserved at higher orders of $\alpha'$, then there should be bulk and boundary actions at higher orders that are invariant under appropriately deformed $\beta$-transformations.

The heterotic string theory has odd parity couplings resulting from the Green-Schwarz anomaly cancellation mechanism \cite{Green:1984sg}. At order $\alpha'$, there is only one odd parity coupling in the bulk, and there is no odd parity coupling on the boundary. Therefore, the $\beta$-transformation of the bulk coupling should produce appropriate deformations at order $\alpha'$ that leave the Gibbons-Hawking boundary term invariant. In this paper, we will perform these calculations in detail.

 \section{The $\beta$-Symmetry at order $\alpha'$ }
 
In this section, we will examine the transformation of the odd parity coupling in the effective action of the heterotic string theory at order $\alpha'$. The effective action of the heterotic string theory has both parity-even and parity-odd parts. The parity-odd part of the effective action at order $\alpha'$ is given by 
\beqa
\bS^{(1)}_O&=&-\frac{2\alpha'}{\kappa^2}\int d^{10} x \sqrt{-G} \,e^{-2\Phi}\left(-\frac{1}{6}H_{\mu\nu\alpha}\Omega^{\mu\nu\alpha}\right).\labell{CS}
\eeqa
where $\Omega$ is the Chern-Simons three-form that arises from the Green-Schwarz anomaly cancellation mechanism \cite{Green:1984sg}. This mechanism requires a non-standard local Lorentz transformation for the $B$-field, given by
\beqa
B_{\mu\nu}\rightarrow B_{\mu\nu}+\alpha' \prt_{[\mu}\Lambda_a{}^b\omega_{\nu]b}{}^a,
\eeqa
where $\Lambda_a{}^b$ is the matrix of the Lorentz transformations and $\omega_{\mu a}{}^b$ is the spin connection. Under this transformation, the 3-form $H_{\mu\nu\alpha}+\alpha' \Omega_{\mu\nu\alpha}$ is invariant, i.e., $H_{\mu\nu\alpha}+\alpha' \Omega_{\mu\nu\alpha}\rightarrow  H_{\mu\nu\alpha}+\alpha' \Omega_{\mu\nu\alpha}$. 
The Chern-Simons three-form $\Omega$ is given by
\beqa
\Omega_{\mu\nu\alpha}&=&\omega_{[\mu a}{}^b\prt_\nu\omega_{\alpha] b}{}^a+\frac{2}{3}\omega_{[\mu a}{}^b\omega_{\nu b}{}^c\omega_{\alpha]c}{}^a,
\eeqa
where $\omega_{\mu a}{}^b=\partial_\mu e_\nu{}^b e^\nu{}_a-\Gamma_{\mu\nu}{}^\rho e_\rho{}^b e^\nu{}_a$, with $e_\mu{}^a$ being  the vielbein, and $e_\mu{}^a e_\nu{}^b\eta_{ab}=G_{\mu\nu}$. The spin connection with subscript indices $\omega_{\mu\nu\alpha}=e_\nu{}^a e_\alpha{}^b\omega_{\mu a b}$ is antisymmetric with respect to its last two indices. 
In the heterotic string theory, by replacing  $H$ with the gauge invariant field strength $\hat{H}=H+\alpha'\Omega$ in the leading order action \reef{baction}, we obtain the odd parity coupling \reef{CS}. 

The $\beta$-transformation of the frame $e_\mu{}^a$ is given by
\beqa
\delta e_\mu{}^a&=& -B_{\mu\alpha}e_{\beta}{}^a\beta^{\alpha\beta}.
\eeqa
This transformation correctly produces the $\beta$-transformation of the metric in \reef{b-trans}. It's worth noting that the $\beta$-transformations change odd parity terms to even parity terms and vice versa. By using the above transformation and the transformations in \reef{b-trans} and applying the constraint \reef{b-cons}, we obtain the transformation of the action \reef{CS} in terms of the curved space fluxes $H_{\mu\nu\alpha}$, $\omega_{\mu\nu\alpha}$, and their covariant derivatives, which can be expressed as\footnote{
The calculations in this paper were performed using the "xAct" package \cite{Nutma:2013zea}. In a previous version of this paper, there was a mistake in calculating the transformation \reef{dCS}. In particular, in that version, the terms without $H$ were missing. This mistake led us to incorrectly conclude that \reef{dCS} is not invariant under the $\beta$-transformations.}
\beqa
\delta \!\!\bS^{(1)}_O&=&-\frac{2\alpha'}{\kappa^2}\int d^{10} x \sqrt{-G} \,e^{-2\Phi}\Big[- \frac{1}{3} \beta^{\alpha \beta } \omega _{\alpha 
}{}^{\gamma \delta } \omega _{\beta }{}^{\epsilon \varepsilon 
} \omega _{\gamma \epsilon }{}^{\mu } 
\omega _{\delta \mu \varepsilon } + 
\frac{1}{6} H_{\gamma \epsilon }{}^{\mu } 
H_{\delta \mu \varepsilon } 
\beta^{\alpha \beta } \omega _{\alpha }{}^{\gamma \delta } 
\omega ^{\epsilon }{}_{\beta }{}^{\varepsilon } \nn\\&&-  
\frac{1}{12} H_{\beta }{}^{\mu \varepsilon 
} H_{\epsilon \mu \varepsilon } \beta 
^{\alpha \beta } \omega _{\alpha }{}^{\gamma \delta } 
\omega ^{\epsilon }{}_{\gamma \delta } + \frac{1}{6} H_{\beta 
\varepsilon }{}^{\mu } H_{\delta \epsilon 
\mu } \beta^{\alpha \beta } \omega _{\alpha }{}^{\gamma \delta } \omega ^{\epsilon }{}_{\gamma 
}{}^{\varepsilon } + \frac{1}{3} H_{\beta \varepsilon 
}{}^{\mu } H_{\gamma \epsilon \mu } \beta^{\alpha \beta } \omega ^{\gamma 
}{}_{\alpha }{}^{\delta } \omega ^{\epsilon }{}_{\delta 
}{}^{\varepsilon } \nn\\&&-  \frac{1}{12} H_{\beta \mu \varepsilon } H_{\gamma \delta \epsilon } \beta^{\alpha \beta } \omega _{\alpha }{}^{\gamma \delta } 
\omega ^{\epsilon \mu \varepsilon } -  
\frac{1}{12} H_{\beta \gamma \epsilon } H_{\delta \mu \varepsilon } \beta^{\alpha \beta } \omega _{\alpha }{}^{\gamma \delta } \omega ^{\epsilon \mu \varepsilon } + \frac{2}{3} \beta^{\alpha 
\beta } \omega _{\alpha }{}^{\gamma \delta } \omega _{\gamma 
}{}^{\epsilon \varepsilon } \omega _{\epsilon \beta 
}{}^{\mu } \omega _{\varepsilon \delta 
\mu }\nn\\&& + \frac{1}{12} H_{\alpha \gamma 
\varepsilon } H_{\beta \epsilon \mu } \beta^{\alpha \beta } \omega ^{\gamma \delta \epsilon } 
\omega ^{\varepsilon }{}_{\delta }{}^{\mu } 
+ \frac{1}{6} H_{\beta \delta }{}^{\varepsilon } H_{\gamma 
\epsilon \varepsilon } \beta^{\alpha \beta } 
\nabla^{\epsilon }\omega ^{\gamma }{}_{\alpha }{}^{\delta } + 
\frac{1}{6} H_{\gamma }{}^{\epsilon \varepsilon } \beta^{\alpha \beta } \omega ^{\gamma }{}_{\alpha }{}^{\delta 
} \nabla_{\varepsilon }H_{\beta \delta \epsilon } \nn\\&&+ 
\frac{1}{12} H_{\alpha \gamma }{}^{\varepsilon } \beta^{
\alpha \beta } \omega ^{\gamma \delta \epsilon } 
\nabla_{\varepsilon }H_{\beta \delta \epsilon } -  \frac{2}{3} 
\beta^{\alpha \beta } \omega _{\alpha }{}^{\gamma 
\delta } \omega _{\gamma }{}^{\epsilon \varepsilon } 
\nabla_{\varepsilon }\omega _{\epsilon \beta \delta }\Big].\labell{dCS}
\eeqa
It is worth noting that the coupling in \reef{CS} is an odd parity term, and its transformation under the $\beta$-transformations is even parity, as expected. 
Using the fundamental requirement that the frame $e_\mu{}^a$ is covariantly constant, i.e.,
\beqa
\nabla_{\mu}e_\nu{}^a\equiv\partial_\mu e_\nu{}^a-\Gamma_{\mu\nu}{}^\rho e_{\rho}{}^a+\omega_{\mu b}{}^a e_{\nu}{}^b=0,
\eeqa
we can rewrite the transformation \reef{dCS} in terms of the flat space fluxes $H_{abc}$, $\omega_{abc}$, and their flat derivatives, for example,
\beqa
\nabla_\mu H_{\nu\alpha\beta}&=&e_\mu{}^a e_\nu{}^b e_\alpha{}^c e_\beta{}^d (D_aH_{bcd}+\omega_{ab}{}^eH_{ecd}+\omega_{ac}{}^eH_{bed}+\omega_{ad}{}^eH_{bce}),
\eeqa
where the flat derivative is $D_a=e^\mu{}_a\partial_\mu$. It is worth noting that the transformation in \reef{dCS} involves only three and four fluxes, regardless of whether we use flat space or curved space fluxes. In fact, it has been observed in \cite{Baron:2022but} that the $\beta$-transformation of the effective action of the heterotic string theory at order $\alpha'$ in the Bergshoeff-de Roo scheme also involves only three and four flat space fluxes. In this paper, we continue our calculations using the curved space tensors.

It is evident that the transformation \reef{dCS} is not zero. However, it is possible that the integrand might be a total derivative or might cancel with terms at order $\alpha'$ that are produced by the transformation of the leading order bulk action \reef{baction} under an appropriate deformation of the $\beta$-transformations at order $\alpha'$.
We consider the most general covariant deformation at order $\alpha'$ for the $\beta$-transformations, given by
\beqa
\delta\Phi&=&\delta\Phi^{(0)}+\alpha' \delta\Phi^{(1)}+\cdots,\nn\\
\delta G_{\alpha\beta}&=&\delta G_{\alpha\beta}^{(0)}+\alpha' \delta G_{\alpha\beta}^{(1)}+\cdots,\nn\\
\delta B_{\alpha\beta}&=&\delta B_{\alpha\beta}^{(0)}+\alpha' \delta B_{\alpha\beta}^{(1)}+\cdots,
\eeqa
where $\delta\Phi^{(0)}$, $\delta G_{\alpha\beta}^{(0)}$, and $\delta B_{\alpha\beta}^{(0)}$ are given in \reef{b-trans}, and $\delta\Phi^{(1)}$, $\delta G_{\alpha\beta}^{(1)}$, and $\delta B_{\alpha\beta}^{(1)}$ are all possible contractions of the fluxes $H_{\mu\nu\alpha}$, $\omega_{\mu\nu\alpha}$, $\nabla_\mu\Phi$, $\beta^{\mu\nu}$, and their covariant derivatives\footnote{Please note that the constant antisymmetric bi-vector $\beta^{\mu\nu}$ is not covariantly constant. In other words, $\nabla_\alpha \beta^{\mu\nu}=0$ does not hold, while $\partial_{\alpha}\beta ^{\mu\nu}= 0$. One could utilize the latter relation to express the covariant derivatives of $\beta^{\mu\nu}$ in terms of the Levi-Civita connection, for example, $ \nabla_\alpha \beta^{\mu\nu}=-2\Gamma^{[\mu}{}_{\alpha\rho}\beta^{\nu]\rho}$. However, in this paper, we refrain from employing such relations.}  at order $\alpha'$. These contractions should include only the linear order of $\beta^{\mu\nu}$.
The transformation of the leading order bulk action \reef{baction} under the above deformed transformation produces the following terms at order $\alpha'$:
\beqa
\alpha'\Delta(\!\!\bS^0)&=&-\frac{2\alpha'}{\kappa^2}\int d^{10} x\sqrt{-G} e^{-2\Phi} \Bigg[ - R^{\alpha \beta } \delta G^{(1)}{}_{\alpha \beta } -  \frac{1}{24} H_{\beta \gamma \delta } H^{\beta \gamma \delta } \delta 
G^{(1)\alpha }{}_{\alpha } + \frac{1}{2} R \delta G^{(1)\alpha 
}{}_{\alpha } \nn\\&&+ \frac{1}{4} H_{\alpha }{}^{\gamma \delta } H_{\beta 
\gamma \delta } \delta G^{(1)\alpha \beta } + \frac{1}{6} H_{\alpha 
\beta \gamma } H^{\alpha \beta \gamma } \delta \Phi^{(1)} - 2 R 
\delta \Phi^{(1)} + 8 \nabla_{\alpha }\Phi \nabla^{\alpha }\delta 
\Phi^{(1)} \nn\\&&+ 2 \delta G^{(1)\beta }{}_{\beta } \nabla_{\alpha }\Phi \nabla^{\alpha }\Phi - 8 \delta \Phi^{(1)} \nabla_{\alpha }\Phi \nabla^{\alpha }\Phi + \nabla_{\beta }\nabla_{\alpha }\delta G^{(1)\alpha 
\beta } -  \nabla_{\beta }\nabla^{\beta }\delta G^{(1)\alpha 
}{}_{\alpha } \nn\\&&- 4 \delta G^{(1)}{}_{\alpha \beta } \nabla^{\alpha }\Phi 
\nabla^{\beta }\Phi -  \frac{1}{2} H_{\alpha \beta \gamma } \nabla^{
\gamma }\delta B^{(1)\alpha \beta }\Bigg].\labell{Ds0}
\eeqa
  Since the transformation \reef{dCS} is even-parity, the deformations $\delta\Phi^{(1)}$, $\delta G_{\alpha\beta}^{(1)}$ must include an even number of $H$, and $\delta B_{\alpha\beta}^{(1)}$ must include an odd number of $H$.
We also include all possible total covariant derivative terms in our calculations. The most general total derivative terms are given by
\beqa
\cJ^{(1)}&=&-\frac{2\alpha'}{\kappa^2}\int d^{10} x\sqrt{-G}\,\nabla_{\alpha}\Big[e^{-2\Phi}I^{(1)\alpha}\Big],
\labell{J}
\eeqa
where the vector $I^{(1)\alpha}$ is all contractions of $\omega$, $H$, $\nabla\Phi$, $\beta$, and their covariant derivatives at order $\alpha'$, which are even-parity and linear in $\beta^{\mu\nu}$.

In order for the action \reef{CS} to be invariant under the deformed $\beta$-transformations, the following relation must be satisfied:
\beqa
\delta\!\!\bS^{(1)}_O+\alpha'\Delta(\!\!\bS^{(0)})+\cJ^{(1)}&\stackrel{}{=}&0.
\labell{cons1}
\eeqa
To solve this equation, one must impose the following Bianchi identities:
\beqa
R_{\alpha[\beta\gamma\delta]}&=&0,\nn\\
\nabla_{[\mu}R_{\alpha\beta]\gamma\delta}&=&0,\labell{bian}\\
\nabla_{[\mu}H_{\alpha\beta\gamma]}&=&0,\nn\\
{[}\nabla,\nabla{]}\mathcal{O}-R\mathcal{O}&=&0,\nn
\eeqa
as well as the constraint \reef{b-cons}.
To impose the Bianchi identities, we write the curvatures, the spin connection, and the covariant derivatives in \reef{cons1} in terms of partial derivatives of the frame $e_\mu{}^a$, and write the field strength $H$ in terms of the $B$-field. In this way, all the Bianchi identities are satisfied automatically, and the constraint \reef{b-cons} can easily be imposed. Then, the equation \reef{cons1} can be written in terms of non-covariant independent terms. If the above relation is correct, then the coefficients of the independent terms should be zero, i.e., there should be total derivative terms and appropriate corrections for the $\beta$-transformations that make the above relation satisfied.

However, some of the coefficients of the total derivative terms and the corrections to the $\beta$-transformations may be related to each other by the Bianchi identities when they are inserted into the constraint \reef{cons1}. Hence, we do not expect the constraint \reef{cons1} to fix all coefficients in the total derivative terms and in the deformations of the $\beta$-transformations. We have found that the equation \reef{cons1} has solutions for many unfixed parameters in the total derivative terms and in the deformations of the $\beta$-transformations. The solution is such that for no values of the remaining parameters, the deformation $\delta G_{\alpha\beta}^{(1)}$ is zero.

Since there is only one odd-parity bulk coupling at order $\alpha'$ with a fixed coefficient, there is no further constraint on the remaining parameters for the closed spacetime manifolds. The remaining parameters are independent of the bulk coupling, so one may set them all to zero. In this way, one finds the following current for total derivative terms:
\beqa
I^{(1)\alpha}&=&\frac{1}{12} H_{\gamma }{}^{\epsilon \varepsilon } H_{\delta 
\epsilon \varepsilon } \beta^{\beta \gamma } \omega_{\beta }{}^{\alpha \delta } -  \frac{1}{3} \beta^{\beta \gamma } \omega_{\beta }{}^{\alpha \delta } 
\omega_{\gamma }{}^{\epsilon \varepsilon } \omega_{\epsilon 
\delta \varepsilon } -  \beta^{\beta \gamma } \omega_{
\beta \delta }{}^{\varepsilon } \omega^{\delta \alpha 
\epsilon } \omega_{\varepsilon \gamma \epsilon }\nn\\&& + 
\frac{1}{3} \nabla_{\gamma }\omega_{\delta }{}^{\alpha }{}_{
\beta } \nabla^{\delta }\beta^{\beta \gamma } + 
\frac{2}{3} \nabla_{\delta }\omega_{\beta }{}^{\alpha 
}{}_{\gamma } \nabla^{\delta }\beta^{\beta \gamma },\labell{I}
\eeqa
and the following deformations for the $\beta$-transformations:
\beqa
\delta\Phi^{(1)}&=&0,\nn\\
\delta G_{\alpha\beta}^{(1)}&=& \frac{1}{3} \beta_{(\alpha }{}^{\gamma } 
\omega_{\beta )}{}^{\delta \epsilon } \omega_{\gamma \delta 
\epsilon } + \frac{1}{3} \nabla_{\gamma }\nabla_{(\alpha 
}\beta_{\beta )}{}^{\gamma }  + \frac{2}{3} \beta_{(\alpha 
}{}^{\gamma } \nabla_{\gamma }\nabla_{\beta )}\Phi  ,\nn\\
\delta B_{\alpha\beta}^{(1)}&=&
\frac{1}{3} H_{[\alpha \delta \epsilon } \beta^{\gamma 
\delta } \omega_{\beta ]\gamma }{}^{\epsilon } -  
\frac{1}{6} H_{\gamma \delta \epsilon } \beta_{[\alpha 
}{}^{\gamma } \omega_{\beta ]}{}^{\delta \epsilon }-  
\frac{1}{3} H_{[\alpha \delta \epsilon } \beta^{\gamma 
\delta } \omega_{\gamma \beta ]}{}^{\epsilon }  + 
\frac{1}{6} H_{[\alpha \delta \epsilon } \beta_{\beta] 
}{}^{\gamma } \omega_{\gamma }{}^{\delta \epsilon },\labell{fff}
\eeqa 
where  $(,)$ is used to symmetrize the indices $\alpha,\beta$, and $[,]$ is used to antisymmetrize $\alpha,\beta$. With the above total derivative terms and deformation of the $\beta$-transformations, the odd parity coupling at order $\alpha'$ is invariant under the $\beta$-transformations. For closed spacetime manifolds, the total derivative terms can be ignored.

Since we are interested in open spacetime manifolds that have a boundary, there are further constraints in addition to the constraint \reef{cons1}. Therefore, we first need to impose these constraints on the parameters, and then the remaining parameters may be set to zero. There is a unit vector $n_{\mu}$ to the boundary that should be invariant under the $\beta$-transformations, i.e., $\delta n_\mu=0$ at all orders of $\alpha'$. Moreover, its length should be invariant. This implies that 
\beqa
n^\mu n^\nu\delta G_{\mu\nu}=0,\labell{unit}
\eeqa
 should hold at all orders of $\alpha'$. 

On the other hand, there are data on the boundary that should be invariant under the $\beta$-transformations. The data at order $\alpha'$ are the values of the massless fields and their first derivatives \cite{Garousi:2021cfc}. In imposing the $O(1,1)$ symmetry on the effective action, since there are no constraints on the massless field in the base space, the constraint \reef{unit} implies that the deformation of the base space metric must be zero. The constraint that the data should be invariant under the $O(1,1)$ transformations also implies that the deformations for other massless fields at order $\alpha'$ should include only the first derivative of the massless fields. 

In the $\beta$-symmetry, however, there are also the constraints \reef{b-cons} as well as the following constraint \cite{Garousi:2022qmk}:
\beqa
\beta^{\mu\nu}n_{\nu}&=&0.\labell{consn}
\eeqa
The constraint mentioned above, which ensures the invariance of the leading-order action described by equation \reef{baction} under the $\beta$-transformations given by equation \reef{b-trans}, was derived in \cite{Garousi:2022qmk}.  These constraints do not allow us to conclude that the deformation of the metric must be zero, and all other deformations must include only the first derivative of $\beta^{\mu\nu}$ and the massless fields. In fact, the data must not be spoiled after using the constraints \reef{b-cons} and \reef{consn}. We clarify this point further in the Conclusion section.

The constraint \reef{unit} dictates that only one of the parameters in the deformations resulting from equation \reef{cons1} is zero, i.e., $\delta G_{\alpha\beta}^{(1)}$ is non-zero even in the presence of a boundary. However, for any values of the parameters, we find that 
\beqa
\delta \Phi^{(1)}=0&;& G^{\mu\nu}\delta G_{\mu\nu}^{(1)}=0,\labell{pG0}
\eeqa
which indicates that the measure $e^{-2\Phi}\sqrt{-G}$ is invariant under the deformed $\beta$-transformations.

In the presence of a boundary, there are similar constraints like \reef{cons1} for the boundary effective action as well. Since there is no odd parity boundary action at order $\alpha'$, the $\beta$-transformation of the leading order boundary term should be canceled with the resulting total derivative terms in \reef{cons1} after using the Stokes' theorem, \ie
  \beqa
\alpha'\Delta(\prt \!\!\bS^{(0)})+\frac{2\alpha'}{\kappa^2}\int d^{9} x\, e^{-2\Phi}\sqrt{|g|}\,n_{\alpha}I^{(1)\alpha}&\stackrel{}{=}&0.
\labell{cons2}
\eeqa 
In the Gaussian normal coordinates, i.e., $x^{\mu}(\sigma^\tmu)=(\sigma^\tmu, z_*)$ where $z_*$ is fixed on the boundary, one has
\beqa
ds^2=G_{\mu\nu}dx^\mu dx^\nu=\pm d^2z+g_{\tmu\tnu}d\sigma^\tmu d\sigma^\tnu\,\,; \,or,\,\,\,\,
G_{\mu\nu}=\left(\matrix{\pm 1&  0&\cr0&g_{\tmu\tnu}&}\right) .
\eeqa
Then using \reef{pG0}, one finds that the measure in the Gibbons-Hawking boundary term in \reef{baction} is invariant under the $\beta$-transformations, i.e.,
\beqa
\delta (e^{-2\Phi}\sqrt{|g|})=e^{-2\Phi}\sqrt{|g|}(-2\delta\Phi^{(1)}+\frac{1}{2}g^{\tmu\tnu}\delta g_{\tmu\tnu})=e^{-2\Phi}\sqrt{|g|}(-2\delta\Phi^{(1)}+\frac{1}{2}G^{\mu\nu}\delta G_{\mu\nu})&=&0.\nn
\eeqa
Hence, we need to study the transformation of the extrinsic curvature of the boundary $K$ under the deformed $\beta$-transformations.

 We consider a spacelike boundary. Using the fact that the normal vector is invariant, we find that the transformation of this curvature is 
\beqa
\delta K&=& -n^\alpha n^\beta(\nabla_\alpha n^\gamma+\nabla^\gamma n_\alpha)\delta G^{(1)}_{\beta\gamma}-n^\alpha\nabla_\beta\delta G^{(1)}_\alpha{}^\beta-\nabla^\beta n^\alpha\delta G^{(1)}_{\alpha\beta}\nn\\&&+\frac{1}{2}n^\alpha\nabla_\alpha\delta G^{(1)}_\beta{}^\beta-\frac{1}{2}n^\alpha n^\beta n^\gamma\nabla_\gamma\delta G^{(1)}_{\alpha\beta}.
\eeqa
When we replace the deformed $\delta G^{(1)}_{\alpha\beta}$ that we have found from the constraints \reef{cons1}, \reef{unit}, and use the constraints \reef{b-cons} and \reef{consn} and their combined constraint $\beta^{\mu\nu}\partial_\alpha n_\mu=0$, we find that the result is zero for any parameter in $\delta G^{(1)}_{\alpha\beta}$. For example, the extrinsic curvature is invariant under the deformation \reef{fff}. However, the total derivative terms \reef{I}  produce some non-zero terms which make the odd parity coupling not invariant under the deformed $\beta$-transformations \reef{fff} in the presence of a boundary.

The boundary constraint \reef{cons2} then dictates that $n_\alpha I^{(1)\alpha}$ must be zero after using the constraints \reef{b-cons} and \reef{consn}. This gives some relations between the parameters in $I^{(1)\alpha}$. Since there is no other constraint that the remaining parameters have to satisfy, they are presumably the dependent parameters that are related to the independent parameters by the Bianchi identities. So we set them all to zero. The result is that there are no total derivative terms, and the deformed $\beta$-transformations are the following:
\beqa
\delta\Phi^{(1)}&=&0,\nn\\
\delta G_{\alpha\beta}^{(1)}&=&\frac{1}{3} \omega_{(\alpha}{}^{\delta \epsilon} \omega_{\gamma\delta \epsilon}\beta_{\beta)}{}^\gamma + \frac{1}{12} H_{(\alpha}{}^{\delta \epsilon} H_{\gamma\delta \epsilon}\beta_{\beta)}{}^\gamma ,\nn\\
\delta B_{\alpha\beta}^{(1)}&=&
\frac{1}{3} H_{[\alpha \delta \epsilon } \beta^{\gamma 
\delta } \omega_{\beta ]\gamma }{}^{\epsilon } -  
\frac{1}{6} H_{\gamma \delta \epsilon } \beta_{[\alpha 
}{}^{\gamma } \omega_{\beta ]}{}^{\delta \epsilon }-  
\frac{1}{3} H_{[\alpha \delta \epsilon } \beta^{\gamma 
\delta } \omega_{\gamma \beta ]}{}^{\epsilon }  + 
\frac{1}{6} H_{[\alpha \delta \epsilon } \beta_{\beta] 
}{}^{\gamma } \omega_{\gamma }{}^{\delta \epsilon },\labell{ff}
\eeqa 
 This ends our illustration that the odd-parity couplings in the heterotic theory at order $\alpha'$ are invariant under the deformed $\beta$-transformation in the presence of a boundary.

\section{Conclusion}

In this paper, we demonstrate the invariance of the odd-parity coupling at order $\alpha'$ in heterotic string theory in the presence of a boundary, as given by \reef{CS}, under the $\beta$-transformations \reef{b-trans} and their deformation at order $\alpha'$ given by \reef{ff}. We consider all possible deformations and total derivative terms at order $\alpha'$ and impose the constraint \reef{b-cons}  on the parameters of the $\beta$-transformations. In particular, we require that the bulk couplings satisfy the $\beta$-constraint \reef{cons1} and the boundary action satisfies the $\beta$-constraint \reef{cons2}. Additionally, in the presence of a boundary, we have the constraints \reef{consn} and \reef{pG0}.
By imposing these constraints, we find that the bulk and boundary actions are invariant under the $\beta$-transformations with no total derivative terms and with the deformations given by \reef{ff}. In our calculation, there are several parameters in the total derivative terms and the deformation that are presumably removable by the Bianchi identity. To simplify the analysis, we set all such parameters to zero.

A similar calculation has been carried out in \cite{Garousi:2023pah} to show the invariance of the circular reduction of bulk and boundary couplings under the $O(1,1)$-group. There are advantages and disadvantages when comparing the two methods. In the $\beta$-symmetry, the advantage is that no Kaluza-Klein reduction is needed, and only massless fields such as the metric, $B$-field, and dilaton need to be considered in the spacetime. The disadvantage is the requirement to impose the constraint \reef{b-cons} and work with a curved spacetime. 
Furthermore, the $\beta$-transformations do not form a group and need to be combined with local Lorentz transformations and $B$-field gauge transformations to satisfy a closed algebra \cite{Baron:2022but}.

On the other hand, the advantage of the $O(1,1)$-symmetry is that there are no constraints in the base space, and the base space can be chosen to be flat. Furthermore, the deformations form the simple $\mathbb{Z}_2$-group, making the calculation in the base space much simpler. However, the disadvantage is that there are more fields in the base space resulting from the Kaluza-Klein reduction than in the original spacetime.

In \cite{Garousi:2021cfc,Garousi:2021yyd}, it was proposed that in the presence of a boundary, there exist data on the boundary that should not be affected by field redefinition or T-duality transformations. Specifically, for the effective action at order $\alpha'$, the boundary data consist of the values of the massless fields and their first derivatives \cite{Garousi:2021cfc}. Therefore, at order $\alpha'$, there should be no field redefinition for the metric, and for other massless fields, any field redefinitions should only involve their first derivatives.

To clarify this point, consider the variation of the leading-order effective action in \reef{Ds0}. The second derivative of the metric perturbation in the third line of \reef{Ds0} produces the first derivative of the perturbation on the boundary using Stokes' theorem. If the metric perturbation includes the first derivative of the massless fields, then the first derivative of the perturbation produces the second derivative of the massless fields on the boundary, which are not known for the effective action at order $\alpha'$. Hence, this perturbation would spoil the boundary data. A similar argument applies to the $\beta$-transformations, which should also preserve the boundary data.

We have observed that the deformation at order $\alpha'$ involves a non-zero $\delta G^{(1)}_{\alpha\beta}$. However, due to the constraints \reef{b-cons} and \reef{consn}, it cannot be concluded that the first derivative of this deformation, which appears on the boundary, produces the second derivative of the massless fields. In fact, the last term in the third line of \reef{Ds0} is zero according to the equation \reef{pG0}. We note that the third term in the third line of \reef{Ds0}, which produces $n_\beta\nabla_\alpha\delta G^{(1)\alpha\beta}$ on the boundary, only generates the first derivative of the massless fields for the deformation in \reef{ff} upon using the constraints \reef{b-cons} and \reef{consn}. Interestingly, the deformation of the $B$-field in \reef{ff} also only involves the first derivative of the massless fields, even without using the constraints \reef{b-cons} and \reef{consn}. Therefore, the deformations \reef{ff} do not spoil the boundary data.

The deformations of the $\beta$-transformations generally depend on the chosen scheme for the effective action. This observation was made in \cite{Garousi:2022qmk} for the parity-even component of the effective action in the heterotic theory at order $\alpha'$. Specifically, the deformation described by equation \reef{ff} corresponds to a particular odd-parity coupling, as given by equation \reef{CS}. This deformation aligns with the findings in \cite{Baron:2022but}, which specifically pertain to the Bergshoeff-de Roo scheme. Such a deformation can be employed to study the even-parity couplings at order $\alpha'^2$ within the Bergshoeff-de Roo scheme \cite{Bergshoeff:1989de}.
In particular, one can examine the transformations of the couplings at order $\alpha'^2$ in this action under the deformed $\beta$-transformations described in equation \reef{b-trans}, in addition to the transformations of the couplings at order $\alpha'$ under the deformed $\beta$-transformations at the same order and the transformation of the leading-order action described by equation \reef{baction} under the $\beta$-transformations at order $\alpha'^2$. These transformations should yield either zero or some total derivative terms. Given that the couplings are known in this scheme, one can utilize it to verify the deformations at order $\alpha'$ and also determine the deformations at order $\alpha'^2$. Performing such calculations would be intriguing in order to identify the deformations of the $\beta$-transformations at order $\alpha'^2$ within the Bergshoeff-de Roo scheme. A similar calculation has been carried out in \cite{Garousi:2023kxw}, where the $O(1,1, \mathbb{Z})$ symmetry was used instead of $\beta$-symmetry, in order to determine the NS-NS couplings at order $\alpha'^2$ in the heterotic theory within the Meissner scheme.

\end{document}